\documentclass[preprint2]{aastex}

\usepackage{epsfig}
\usepackage{amsmath}

\slugcomment{Submitted to ApJ Letters}

\shorttitle{Salmonson}
\shortauthors{Kinematics of Gamma-ray Bursts}

\begin{document}


\title{On the Kinematic Origin of the Luminosity-Pulse Lag Relationship in Gamma-ray Bursts}


\author{Jay D. Salmonson}
\affil{Lawrence Livermore National Laboratory, Livermore, CA 94550}



\begin{abstract}

This paper presents an interpretation based on gamma-ray burst source
kinematics for the relationship found by Norris et al.~between
peak luminosity and energy-dependent pulse lag.  I argue that the
correlation should instead be between {\it number} luminosity and
pulse lag.  This interpretation improves the least-squares fit of this
correlation for the known bursts by 25 percent or more.  It also
suggests a distance estimation scheme.  I propose that this
relationship is due to the variation in line-of-sight velocity among
bursts.  This interpretation allows one to speculate on the range of
gamma-ray burst expansion velocities or the size of their jet opening
angles.
\end{abstract}




\section{Introduction}

It has recently been reported by \citet{nmb00,nmb00b} that there is a
relationship between the peak luminosity of gamma-ray bursts (GRB) and
the pulse time lag between BATSE energy channels.  Specifically, they
find that the peak luminosity $L_{pk}$ is anticorrelated with the duration
of the lag, $\Delta t$, according to $L_{pk} \approx 1.3 \times (\Delta
t/0.01 \text{ sec})^{-1.15} \times 10^{53}$ ergs/sec .  This intriguing
relationship is useful because it probes the properties of an ensemble
of bursts and thereby can provide clues of the global dynamics of
GRBs.  It also has the potential for distance estimation to GRBs.

Several workers have attempted to explain the origin of the pulse lag,
or so called pulse paradigm \citep{nnb+96}.  \citet{dm98} were able to
produce a pulse lag by modeling the burst as a series of internal
shocks, but obtained timescales an order of magnitude too large.
However, as \citet{wf00} point out, the lag timescale is much longer
than the synchrotron cooling time.  \citet{pm98} modeled the
kinematics of internal shocks and found no evidence of pulse peak lag.
As of yet, the source of the pulse paradigm remains a mystery.  This
paper does not attempt to explain the origin of the pulse lag, but
assumes that it derives from a process common to all bursts and has
some proper decay timescale $\Delta t'$ in the reference frame of the
emitter.

In this paper I will present an interpretation of the luminosity-lag
relationship as being due to relativistic motion of the emitting
region toward the observer.  Thus it is a purely kinematic effect.

\section{Photon Number Luminosity and Cosmology}

In \citet{nmb00} a peak $\gamma$-ray luminosity $L_{pk}$ was defined
from the observed peak {\it number} flux $F_{pk}$, a luminosity
distance $D_L$ derived from the observed redshift $z$ and a given
cosmology, and a mean emitted photon energy $\overline{\epsilon}$,
constant for all bursts:
\begin{equation}
L_{pk} \equiv 4\pi \overline{\epsilon} F_{pk} D_L^2 ~.
\label{lumeqn}
\end{equation}
However, one does not know the intrinsic energy of the photons
emanating from the pulses.  Therefore multiplication of all GRB fluxes
by an average photon energy $\overline{\epsilon}$, while useful to
estimate GRB peak luminosites, obfuscates the underlying relationships
in the dynamics of GRBs.  In other words, photon number flux $F_{pk}$
is the observed quantity and thus photon number luminosity $N_{pk}$
can be the only calculable luminosity.

The luminosity distance $D_L$ is defined so that redshift of photons
is accounted for.  The photon number luminosity does not depend on
photon energy redshift and thus is given by
\begin{equation}
N_{pk} = \frac{4 \pi F_{pk} D_L^2}{(1+z)} ~. 
\label{numlumeqn}
\end{equation}

\begin{figure}[tb]
%\centering
\epsfig{figure=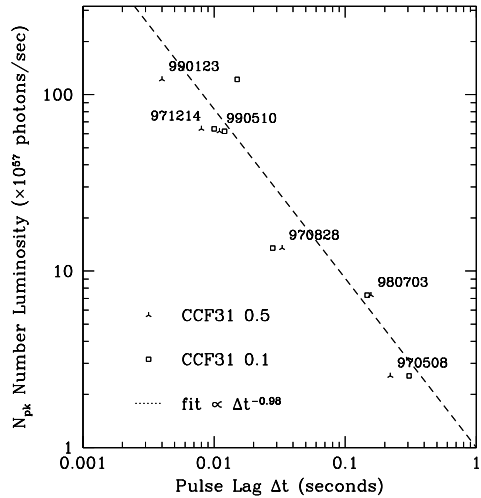, width = 7 cm}
\caption{Peak photon number luminosity $N_{pk}$ versus spectral pulse
lag for six bursts with known redshifts.  Spectral cross-correlation
function lag between BATSE channels 3 and 1 (CCF31) for regions down
to 0.5 and 0.1 of peak intensity were obtained from \citet{nmb00}.
The line of best fit for 0.1 (squares) is $\propto \Delta
t^{-0.98}$. }
\label{numberluminosity}
\end{figure}

In Figure \ref{numberluminosity} is shown the peak number luminosity
$N_{pk}$ versus spectral lag $\Delta t$ for six bursts with known
redshifts.  I fit the data by minimizing $Q^2 \equiv \Sigma
(\log(N_{pk,\text{model}}) - \log(N_{pk,\text{data}}))^2$ where
$N_{pk,\text{model}} \equiv A (\Delta t) ^{-p}$. The best fit for
lags CCF31 0.1 (see Fig. \ref{numberluminosity}) is 
\begin{equation}
N_{pk} = 8.6 \times 10^{56} \Delta t^{-0.98} \text{ photons sec$^{-1}$}
\end{equation}
with $Q^2 = 1.38$.  By comparison, a similar fit of $L_{pk}$ from Eqn
(\ref{lumeqn}) versus $\Delta t$ reproduces the \citet{nmb00} result
$L_{pk} \propto \Delta t^{-1.15}$ with $Q^2 = 1.73$.  For CCF31 0.5,
$N_{pk} \propto \Delta t^{-0.89}$ with $Q^2 = 0.45$ contrasts with
$L_{pk} \propto \Delta t^{-1.03}$ with $Q^2 = 0.79$.  With six
datapoints, these are small number statistics to be sure.  However,
the improvement in correlation $N_{pk}$ vs. $\Delta t$ over that of
$L_{pk}$ vs. $\Delta t$ via reductions of $Q^2$ by 25\% or greater is
clear.

\section{Relativistic Kinematics}

The relation $N_{pk} \propto \Delta t^{-0.98}$, where 0.98 is almost
1, is a strikingly simple statistical property for an ensemble of GRBs
to obey.  In fact this relationship suggests a simple explanation.  I
propose that this proportionality is a result of variations in
line-of-sight relativistic velocities of GRBs.

For a burst expanding with Lorentz factor $\gamma \equiv 1/\sqrt{1 -
\beta^2}$, $\beta \equiv v/c$ and at redshift $z$, the observed
number luminosity $N_{obs}$ (photons sec$^{-1}$) varies as
\begin{equation}
N_{obs} = \frac{\gamma N' }{ (1+z)}
\label{Nobseqn}
\end{equation}
where $N'$ is the proper isotropic number luminosity for a
non-expanding source.  This expression is valid for emission that
appears isotropic to the observer (i.e. if the emission is in the form
of a jet, then the opening angle $\theta_0 > 1/\gamma$, where
$1/\gamma$ is the relativistic beaming angle).  Also, as mentioned
previously, if one assumes that the spectral pulse lag is due to some
proper decay timescale $\Delta t'$, possibly due to cooling or
deceleration, then we observe a lag
\begin{equation}
\Delta t_{obs} = \gamma (1 - \beta \cos \theta)(1+z) \Delta t' \simeq \frac{(1 + z) \Delta t'}{2 \gamma} 
\label{dtobseqn}
\end{equation}
for $\gamma \gg 1$, where the angular dependence is removed by the
strong effect of attenuation of number flux $F$ received from emission
moving at inclination angles greater than $1/\gamma$; $F \sim (\gamma
(1 - \beta \cos \theta) (1+z))^{-3}$. Thus one can focus one's
attention only on emitters moving directly toward the observer.
Combining these then yields the inverse relationship
\begin{equation}
N_{obs} \propto \Delta t_{obs}^{-1} ~.
\label{nteqn}
\end{equation}
This relationship is redshift independent and thus is valid for
luminosity $N_{pk} = (1+z) N_{obs}$ and spectral lag $\Delta t =
\Delta t_{obs}/(1+z)$ as would be observed local to the burst.  Thus
we have an explanation of the relation shown in Figure
\ref{numberluminosity}.  I propose that the variety in the spectral
lags and number luminosities of bursts depend solely upon the
relativistic motion of the emitter along the line of sight.  Certainly
one expects variation among bursts to cause deviations from this
relationship, however the current data suggests this effect to be
dominant.

There are two possibilities for the nature of this variation (see
Section \ref{sectionbeaming}).  First, we are observing an intrinsic
range of burst expansion speeds, ranging from fast bursts having
little or no spectral lag to slower bursts having longer lags.  The
second possibility is that emitting material is in the form of a jet
with some angular dependence on the emitter's Lorentz factor
$\gamma(\theta)$ where the Lorentz factor would have a peak at the
centroid of the jet $\gamma(\theta = 0) = \gamma_{max}$ and
monotonically decrease on some angular scale greater than $1/\gamma$
for increasing inclination angles $\theta$.  In either case we observe
only the emission dominated by those emitters moving with trajectories
inclined within an angle $1/\gamma$ from our line-of-sight.

\section{ Redshift Estimation }

The above discussion allows the estimation of redshift $z$ from the
observed peak number flux $F_{pk}$ and the observed spectral lag
$\Delta \tau_{obs}$.  \citet{frr00} also estimate redshifts using GRB
variability to indicate luminosity.  From Eqn (\ref{numlumeqn}) one has
\begin{equation}
N_{pk} = 4 \pi F_{pk} (1 + z)^3 D_A^2
\label{numlumeqn2}
\end{equation}
where $D_A(z)$ is the angular diameter distance \citep[e.g.][]{hogg99}
which is a function of the cosmology and redshift $z$, but does not
vary much over the interval $1 \lesssim z \lesssim 3$.  For example,
here I follow \citet{nmb00} and choose ($\Omega_M$,$\Omega_\Lambda$) =
(0.3,0.7), for which $D_A \approx 0.4 D_H$ to within 10\% over this
range of $z$, where $D_H \equiv c/H_0$ is the Hubble Distance (choose
$H_0 = 65$ km sec$^{-1}$ Mpc$^{-1}$).  Also, an observed spectral lag
$\Delta t_{obs}$ varies from the local spectral lag $\Delta t$ by
\begin{equation}
\Delta t = \frac{\Delta t_{obs}}{( 1 + z )} ~.
\end{equation}

One can put these components into the $N_{pk}(\Delta t)$ curve of bursts
with known redshifts
\begin{equation}
N_{pk} = \frac{A}{ \Delta t} ~,
\label{ntrelation}
\end{equation}
where $A = 8.6 \times 10^{56}$ photons as taken from the fit in Figure
\ref{numberluminosity}.  Thus one can estimate a redshift for a given
observed peak number flux and spectral lag
\begin{equation}
z \simeq \sqrt{\frac{A}{4 \pi F_{pk} D_A^2 \Delta t_{obs}}} - 1 
\simeq \frac{1.5}{ \sqrt{F_{pk} \Delta t_{obs}}} - 1
\label{zeqn}
\end{equation}
where flux $F_{pk}$ is measured in photons cm$^{-2}$ sec$^{-1}$ and
$\Delta t_{obs}$ is in seconds.  It is worth noting that a well
determined cosmology gives a well determined $D_A(z)$ and thus a
refined redshift $z$.  Conversely, many independently determined
redshifts might give cosmological information.  However, the primary
source of error in this relation is not the uncertainty in $D_A(z)$,
but in the deviations of burst luminosities and lags from the relation
of Eqn (\ref{ntrelation}) as can be seen in Figure
\ref{numberluminosity}.

\section{ Observed Peak Pulse Energy }

In order for there to be a lag $\Delta t$ of the arrival of a pulse
over an energy range fixed with respect to the observer (e.g. the
range between BATSE channels 1 and 3) that varies inversely with
Lorentz factor (Eqn \ref{dtobseqn}), the peak energy of the pulse must
decay exponentially as
\begin{equation}
E_{pk} \simeq E'_{pk,0} \gamma \exp{\biggl( - \frac{t \gamma}{\tau_{pk}}\biggr)}
\label{Epkeqn}
\end{equation} 
where $E'_{pk,0}$ is the peak energy in the emitter frame, $t$ is observer
time, $\gamma$ is the Lorentz factor of the emitter and $\tau_{pk}$ is
the proper decay timescale in the emitter frame.  Thus we infer an
exponential decay law for the pulse peak energy.

It is important to note that this exponential evolution can be very
rapid ($\lesssim 10$ msec) \citep{nnb+96,nmb00} and is distinct from the
evolution of the pulse envelope (the so-called FRED) that has been
studied by many authors \citep[e.g.][]{rs00,lk96,nnb+96, fb95}.

One outstanding question in the study of GRB pulses is the apparent
invariance among bursts of the peak energy $E_{mx}$ of the $E^2 N_{E}$
flux spectrum ($E_{mx} \sim$ a few $\times 100$ keV) where $N_{E}$ is
number flux per photon energy.  One might expect this value to vary
with burst luminosity, but it remains roughly constant (to a factor of
a few) over a wide variety of (long) GRBs.

If the pulse photon number flux $N_E$ (which, for demonstration, we
assume to be peaked at $E_{pk}$) increases as the peak photon energy
$E_{pk}$ of Eqn (\ref{Epkeqn}) decreases, then there will be a maximum
in $E^2_{pk} N_E$.  This may happen as follows: the nascent pulse
emitter is extremely hot and compact.  As it evolves, it cools
($E_{pk}$ drops) and it expands or becomes progressively more
optically thin, releasing an ever increasing photon rate $N_E$. This
is analogous to early-time supernova lightcurves \citep{ewwp94}.  In
this picture, uncovering the flux dependence $N_E(t,\gamma)$ will give
insight into why $E_{mx} \approx$ constant.  For instance $N_E \propto
t^{\ln \gamma} = \gamma^{\ln t}$ with Eqn (\ref{Epkeqn}) demonstrates
this behavior.

\section{ Discussion } \label{sectionbeaming}

In this letter I have argued that the correlation of \citet{nmb00} is
between number luminosity and spectral lag and I interpret this as
being due to the variety, among bursts, of relativistic velocities at
which emitting regions move toward the observer.  Suppositions as to
the physical emission process have been deliberately avoided in order
to highlight this purely kinematic effect.

There are two possible scenarios by which this might happen.  The
first is simply that the variation is due to the variation of
expansion Lorentz factor among bursts.  From Eqn (\ref{dtobseqn}),
this implies that the ratio of maximum to minimum spectral lags is the
same as the ratio of maximum to minumum Lorentz factors
\begin{equation}
\frac{\gamma_{max}}{\gamma_{min}} = {\frac{\Delta t_{max}}{\Delta
 t_{min}}} \gtrsim  100 
\end{equation}
where observed lags are seen to range over roughly two orders of
magnitude.  Thus if the fastest bursts have $\gamma_{max} \sim 100$,
then the minimum bursts would have $\gamma_{min} \sim 1$ and thus may
be only mildly relativistic.

The second possibility is that GRB ejecta is directed in a jet such
that the Lorentz factor $\gamma(\theta)$ has some maximum
$\gamma(\theta = 0) = \gamma_{max}$ and monotonically decreases with
increasing angle. Thus the function $\gamma(\theta)$ will determine
the relative numbers of bursts observed with given spectral lags.
\citet{band97} has noted that the distribution of bursts is strongly
peaked at small lags indicating that the Lorentz factor of emitting
ejecta, $\gamma(\theta < \theta_0/2) \sim \gamma_{max}$, is basically
constant over the jet opening angle $\theta_0$, with a fairly narrow
edge region $\theta > \theta_0$ of decreasing $\gamma$.  Thus let us
estimate a characteristic opening angle $\theta_0$ by assuming that
$\gamma = \gamma_{max}$ is constant for $\theta < \theta_0/2$ and
decreases to order unity within a narrow boundary or ``edge'' region
$0 < \theta - \theta_0/2 < 1/\gamma_{max}$.  Thus the solid angle
subtended by ``face'' bursts (i.e. observed at inclinations $\theta <
\theta_0/2$) is $\pi (\theta_0/2)^2$, and that for ``edge'' bursts is
$\pi \theta_0/\gamma_{max}$ and the ratio is:
\begin{equation}
\frac{\text{``edge'' bursts}}{\text{``face'' bursts}} =
\frac{4}{\gamma_{max} \theta_0} ~.
\end{equation}
From Fig. 3 of \citet{nmb00}, one can estimate this ratio, by defining
short lags $< 0.1$ sec as ``face'' and the longer lags as ``edge''
bursts, to be $\sim 2/5$.  Assuming $\gamma_{max} = 100$, then this gives
$5^o < \theta_0 < 10^o$.  With more statistics of the distribution of
bursts along the curve of Eqn. (\ref{nteqn}) one can fit the function
$\gamma(\theta)$ much more precisely.

This work was performed under the auspices of the U.S. Department of
Energy by University of California Lawrence Livermore National
Laboratory under contract W-7405-ENG-48.\\

\clearpage

\end{document}